\documentclass[russian,aps]{revtex4}
\usepackage{babel}
\usepackage{graphicx}

\begin{document}

\begin{Large}

\title{SOLITON-LIKE EXCITATION IN LARGE-$\alpha$ QED}

\author{I.D.Feranchuk and S.I.Feranchuk}

\address{Belarusian State University,4, Nezavisimosti Ave., 220050 Minsk, Republic of Belarus \\
E-mail: fer@open.by\\
Tel.: (375 172) 277 617; Fax: (375 172) 002 353}

\begin{abstract}

\noindent \textbf{Abstract}\\

The nonperturbative analysis of the one-particle excitation of the electron-positron field is made in the paper.
The standard form of quantum electrodynamics (QED) is used but the coupling constant $\alpha_0$ is supposed to be
of a large value ($\alpha_0 \gg 1$). It is shown that in this case the quasi-particle excitation can be produced
together with the non-zero scalar component of the electromagnetic field. Self-consistent equations for spatially
localized charge distribution coupled with an electromagnetic field are derived. Soliton-like solution with a
nonzero charge for these equations are calculated numerically. The solution proves to be unique if the coupling
constant is fixed. It leads to the condition of charge quantization if the non-overlapping $n$-soliton states are
considered. It is also proved that the dispersion law of the soliton-like excitation is consistent with Lorentz
invariance of the QED equations.
\end{abstract}

\maketitle


\newpage

\section{Introduction}

The problem of quantum electrodynamics (QED) with strong a coupling between electron-positron and electromagnetic
fields (the so-called "large-$\alpha$ QED") was developed by many authors (for example, \cite{large} and citation
therein). QED is a physical system with well defined Hamiltonian, so the approbation of various methods for this
theory is of great interest for non-perturbative analysis of actual problems of quantum chromodynamics (for
example, \cite{Kleinert},\cite{QCD} and citation therein).

One of the main approaches in this field is related to the Schwinger-Dyson equation (SD). the critical value of
the coupling constant $\alpha_c \sim 1 $ separating the regions of weak and strong coupling was discovered for
this equation \cite{SD}. The existence of the critical point $\alpha_c$ could affect some observed physical
phenomena as some authors considered \cite{SD1}. Actually the SD equation defines the self-energy part of the
fermion propagator and results from the accurate QED equations if the corrections for the vertex function and the
Green's function of the electromagnetic field are not taken into account.  This corresponds to one of the
variants of "quenched QED" when the truncated Fock basis is used for the variational perturbation theory
\cite{Mors}. A detailed analysis of nonperturbative approaches in large-$\alpha$ QED has been recently discussed
in paper \cite{large}.

It is important to pay attention to the fact that the mathematical structure of the interaction operator in QED
is similar in many respects to the operator of the interaction between electron and quantum field of optical
phonons in the ionic crystal, the so-called problem of a large-radius "polaron" described by the Fr\"ohlich
Hamiltonian \cite{Frohlich}. This situation was recently used by authors of \cite{Alex} for nonperturbative mass
renormalization in the framework of "quenched QED". They made the variational estimation of the functional
integral for QED on the basis of Feynman's approach to the "polaron" problem \cite{Feynmanpol}.

The main effect of the strong interaction between the electron and quantum field is defined by formation of the
soliton-like ("polaronic") self-localized state of the particle coupled with the classical components which are
selected from the quantum field by the particle itself. It was first discovered by Pekar \cite{Pekar} and
investigated rigorously in the papers \cite{Bogol},\cite{Tyablikov}. It was also proved that the soliton-like
excitation (SLE) corresponds to the leading term in the series of $\alpha^{-1}$ for the Fr\"ohlich Hamiltonian
\cite{Bogol}. Later a lot of different forms of the quasi-particles arising as a result of a "polaronic" effect
were described for many concrete physical systems.

Then the following question is of interest: could some SLE exist in the framework of "large-$\alpha$ QED"? It is
important to mention that the answer to this question cannot be found on the basis of SD equation. The matter is
that the integral equation for the self-energy function was introduced for the "polaron" problem by Pines
\cite{Pines} and it is analogous to SD equation for QED. However, it is well known \cite{polaron}, that Pines
equation does not allow one to describe  the strong coupling limit for the "polaron" problem correctly because
the bound state of the electron and phonons is not included in this equation. It appeares that the essential
reconstruction of the wave function both for electron and phonon field should be taken into account in zeroth
approximation directly if the series in the power of $\alpha^{-1}$ is used for describing of SLE
\cite{Bogol},\cite{Tyablikov}.  If the same problem is considered on the basis of the variational estimation for
functional integrals some model system qualitatively different from noninteracting particles should also be used
for calculating the action in zeroth approximation directly \cite{Feynmanpol}. Particularly, the harmonic
interaction between electron and some fictitious particle was used for the "polaron" model, that the coupling
constant and the mass of this particle being considered as the variational parameters \cite{Feynmanpol}.

In this paper it is shown for the first time that SLE can also be  found for QED if the original coupling
constant for the interaction between electromagnetic and electron-positron fields is supposed to be large
$\alpha_0 \gg 1$. As distinct from "polaron" the self-localized SLE for QED is the quasi-particle excitation of
the electron-positron field corresponding to a spatial charge distribution with a non-zero integral charge
coupled with a classical scalar component of the electromagnetic field separated by the same charge density (see
also \cite{electron}). In the terms of "quenched QED" it corresponds to the truncated Fock basis which consists
of the linear superposition of one-particle excitation of the electron-positron field, non-zero scalar field and
vacuum state for the transversal photons of the electromagnetic field.

The truncation of Fock space used in this paper  differs greatly from other non-perturbative approaches (see, for
example, \cite{large} and references therein) because of the inclusion of a strong localized scalar field.
However, it should be noticed that a similar variant of the quenched QED was recently considered in the paper
\cite{Hainzl} on the basis of mean-field approximation. Some general theorems were proved in this paper but the
calculation of some concrete characteristics of the system was not given.

It is also shown in the paper that SLE with a non-zero charge corresponds to the minimum of the energy for
one-paricle excitation of the system. The characteristics of SLE are defined by self-consistent equations for
electron-positron wave functions and a scalar field. The equations are derived on the basis of an operator method
(OM) introduced earlier for non-perturbative description of quantum systems ( for example, \cite{OM} and
references therein). These results are described in $\S$ 2.

The solution of these nonlinear self-consistent equations is considered in $\S$ 3 on the basis of the numerical
method developed earlier for the "polaron" problem \cite{Krylov}. It is shown that a non-trivial solution of
these equations  exists for a unique value of SLE integral charge. It can be considered as the condition of a
charge quantization similar to that which appeared due to Dirac's monopole \cite{monopol} .

The dependence of SLE energy $E(\vec P)$ on its total momentum $\vec P$ is analyzed  in  $\S$ 3. It is well known
\cite{Lifshitz} that this dispersion law for any one-particle excitation should be of a relativistic form because
of Lorentz-invariance of QED. The problem of the relation between the localization of electron and translational
invariance of the whole system also appeared for "polaron". It was rigorously solved by authors
\cite{Bogol},\cite{Gross}. They proved that the "polaron" state vector did not actually break the translational
symmetry of the system because the concrete point of its localization could be arbitrary in the space
\cite{quasi}.  As is known, the dispersion law is rather complicated for the "polaron" problem and it is very
difficult to find  its general form analytically because the variables describing the internal dynamics of the
system are not separated from the translational variables \cite{Gross}.

The calculation of $E(\vec P)$ for QED ($\S$ 3) showed that SLE localization did not contradict
Lorentz-invariance of the whole system. It is also important that SLE internal degrees of freedom are separated
from the translational motion and as a result its  dispersion law $E(\vec P)$ corresponds exactly to conventional
dynamics of a relativistic free particle.

\section{Self-consistent equations for one-particle excitation in QED}

Let's consider an exact Hamiltonian QED in the Coulomb gauge \cite{Heitler}, \cite{Hainzl}. This representation
is most convenient for us because it extracts the electrostatic interaction which, by definition, makes the main
contribution to the vacuum polarization in the strong coupling limit:

\begin{eqnarray}
\label{Eqn18} \hat H = \int d \vec{r}\{ \hat \psi^* (\vec{r}) [ \vec \alpha (\vec{p} + e_0 \hat{ \vec{ A} }
(\vec{r})) + \beta m_0] \hat \psi (\vec{r}) +  e_0\hat \varphi (\vec{r}) \hat \rho (\vec{r}) - \frac{1}{2} (
\vec{\nabla}\hat \varphi
(\vec{r}))^2\} + \nonumber\\
\sum_{\vec k \lambda} \omega(\vec k) \hat n_{\vec k \lambda};\nonumber\\
\hat \rho (\vec{r}) = \frac{1}{2} [\hat \psi^* (\vec{r})\hat \psi (\vec{r}) - \hat \psi (\vec{r})\hat \psi^*
(\vec{r})].
\end{eqnarray}

We suppose here that the field operators are given in the Schr\"odinger  representation, the spinor components of
the electron-positron operators being defined in the standard way \cite{Heitler}

\begin{eqnarray}
\label{Eqn19} \hat \psi_{\nu} (\vec{r}) = \sum_{s} \int \frac{d\vec{p}}{(2\pi)^{3/2}} \{a_{\vec{p}s}
u_{\vec{p}s\nu} e^{i\vec{p}\vec{r}} +
b^+_{\vec{p}s} v_{-\vec{p}-s\nu} e^{-i\vec{p}\vec{r}}\};\nonumber\\
\hat \psi^*_{\nu} (\vec{r}) = \sum_{s} \int \frac{d\vec{p}}{(2\pi)^{3/2}} \{a^+_{\vec{p}s} u^*_{\vec{p}s\nu}
e^{-i\vec{p}\vec{r}} + b_{\vec{p}s} v^*_{-\vec{p}-s\nu} e^{i\vec{p}\vec{r}}\}.
\end{eqnarray}

In these formulas $\vec{\alpha}, \beta$ are Dirac matrixes; $u_{\vec{p}s\nu}$ and $v_{\vec{p}s\nu}$ are the
components of the bispinors corresponding to the solutions of  Dirac equation for the free "bare" electron and
positron with the momentum $\vec{p}$ and spin s; $a_{\vec{p}s}(a^+_{\vec{p}s})$ and
$b_{\vec{p}s}(b^+_{\vec{p}s})$ are  the annihilation (creation) operators for the "bare" electrons and positrons
in the corresponding states.  The field operator $\hat{ \vec{A}}(\vec{r})$ and the operator of the photon number
$\hat n_{\vec k \lambda}$ are related to the transversal electromagnetic field and their explicit form will be
written below. It should be remembered that the parameter  $e_0$ was introduced as a positively defined value.

We should also discuss the dependence of the QED hamiltonian on the operator of the scalar field

\begin{equation}
\label{19a} \hat \varphi (\vec{r}) = \sqrt{4\pi}\int d\vec{k} \hat \varphi_{\vec{k}}e^{i\vec{k}\vec{r}} .
\end{equation}

It is known\cite{Heitler}, that these operators can be excluded from the Hamiltonian in the Coulomb gauge. For
that purpose one should use the solution of the operator equations of motion for   $\hat \varphi_{\vec{k}}$
assuming that the "bare" electrons are point-like particles and  "self-action" is equivalent to the substitution
of the initial mass for the renormalized one. As a result the terms with scalar fields in the Hamiltonian are
reduced to the Coulomb interaction between the charged particles. However, this transformation of the hamiltonian
(\ref{Eqn18}) cannot be used in this paper because just the dynamics of the mass renormalization is the subject
under investigation.

There is another problem connected with a negative sign of the term corresponding to the self-energy of the
scalar field. If the non-relativistic problems were considered then the operator of the particle kinetic energy
would be positively defined and the negative operator with the square-law dependence on  $\hat\varphi (\vec{r})$
would lead to the "fall on the center" \cite{Landau} as the energy minimum would be reached at an infinitely
large field amplitude. However, if the relativistic fermionic field is considered then the operator of the free
particle energy (the first term in formula (\ref{Eqn18})) is not positively defined. Besides, the states of the
system  with the negative energy are filled. Therefore, the stable state of the system corresponds to the energy
extremum (!) (the minimum one for electron and the maximum one for positron excited states). It can be reached at
the finite value of the field amplitude (see below). The same reasons enable one to successfully use the states
with indefinite metric \cite{Akhiezer} in QED although it leads to some difficulties in the non-relativistic
quantum mechanics.

According to our main assumption about the large value of the initial coupling constant  $e_0$ we are to realize
the nonperturbative description of the excited state which is the nearest to the vacuum state of the system. The
basic method for the nonperturbative estimation of the energy is the variational approach with some trial state
vector $|\Phi_1 >$ for the approximate description of the one-particle excitation. The qualitative properties of
the self-consistent excitation in the strong coupling limit \cite{Pekar} show that such trial vector should
correspond to the general form of the wave packet formed by the one-particle excitations of the "bare"
electron-positron field. Besides, the effect of polarization and the appearance of the electrostatic field
$\varphi (\vec{r})$ should be taken into account, so we consider  $|\Phi_1>$ to be the eigenvector for the
operator of the scalar field. Now, let's introduce the following trial vector depending on the set of variational
classical functions $U_{\vec q s}; V_{\vec q s};\varphi (\vec r)$ for the approximate description of the
quasi-particle excited state of the system:

\begin{eqnarray}
\label{Eqn20} |\Phi_1 > \simeq |\Phi^{(0)}_1(U_{\vec q s}; V_{\vec q s};\varphi (\vec r) )> = \int d \vec{q} \{
U_{\vec{q}s} a^+_{\vec{q} s} + V_{\vec{q}s} b^+_{\vec{q} s} \} | 0;
0;\varphi(\vec r)>; \nonumber\\
\hat \varphi(\vec r)| 0; 0;\varphi(\vec r)> =\varphi(\vec r)| 0; 0;\varphi(\vec r)>; \quad a_{\vec{q} s}| 0;
0;\varphi(\vec r)> = b_{\vec{q} s}| 0; 0;\varphi(\vec r)> = 0.
\end{eqnarray}

The ground state of the system is   $|\Phi_0> = |0;0;0>$, if we use the same notation. It corresponds to the
vacuum of both interacting fields.

Firstly, let's consider the excitation with the zero total momentum. Then the constructed trial vector should
satisfy the normalized conditions resulting from the definition of the total momentum $\vec{P}$ and the observed
charge $e$ of the "physical" particle:

\begin{eqnarray}
\label{Eqn21} <\Phi^{(0)}_1|\hat{ \vec {P}}|\Phi^{(0)}_1> = \sum_{s}\ d \vec{q} \vec{q} [|U_{\vec{q}s}|^2 +
|V_{\vec{q}s}|^2] = \vec{P} = 0;\nonumber\\
\sum_{s}\ d \vec{q} [|U_{qs}|^2 + |V_{qs}|^2] = 1.
\end{eqnarray}

\begin{eqnarray}
\label{Eqn21a} <\Phi^{(0)}_1|\hat Q|\Phi^{(0)}_1> = e_0 \sum_{s}\ d \vec{q} [|V_{qs}|^2 - |U_{qs}|^2] = e.
\end{eqnarray}

The condition (\ref{Eqn21}) requires that the functions $U_{qs}$ and $ V_{qs}$ should depend on the modulus of
the vector $\vec{q}$ only. Besides, one should take into account that the trial vector $|\Phi^{(0)}_1>$ is not
the accurate eigenvector of the exact integrals of motion $\hat Q$ and $\hat{ \vec {P}}$ as it represents the
accurate eigenvector of the hamiltonian $|\Phi_1>$ only approximately. Therefore, in the considered zero
approximation the conservation laws for momentum and charge can be satisfied only on  average, and this leads to
the above written normalized conditions. Generally, equation  (\ref{Eqn21a}) should not be considered as the
additional condition for the variational parameters but as the definition of the observed charge of the
"physical" particle at the given value of the initial charge of the "bare" particle. Therefore the sign of the
observed charge is not fixed a priori,the same was the case for the qualitative estimation in \cite{electron}.
Calculating the sequential approximations to the exact state vector $|\Phi_1>$ (see $\S 4$) should restore the
accurate integral of motion as well. An analogous problem appears in the "polaron" theory when the momentum
conservation law was taken into account for the case of the strong coupling (for example, \cite{Bogol},
\cite{Gross}).

In this respect the variational approach differs greatly from the perturbation theory where the zero
approximation for one-particle state is described by one of the following state vectors:

\begin{eqnarray}
\label{Eqn21b} |\Phi_1> \simeq |\Phi^{(PT)}_1e> = a^+_{\vec{P} s} | 0; 0; 0 >; \quad |\Phi_1> \simeq
|\Phi^{(PT)}_1p> = b^+_{\vec{P} s} | 0; 0; 0 >.
\end{eqnarray}

These vectors don't depend on any parameters and are eigenvectors of the momentum and charge operators. But they
correspond to one-particle excitations determined by the charge $e_0$ of the "bare" electron and the field
$\varphi (\vec r ) = 0$. We suppose that the introduction of the variational parameters into the wave function of
the zero approximation will enable us to take into account the vacuum polarization, but the variational approach
in this form allows to conserve the exact integral of motion only on average. This problem will be considered in
more detail below ($\S$ 4).

So, the following variational estimation for the energy $E_1 (\vec{P}=0)$ of the state corresponding to the
"physical" quasi-particle excitation of the whole system is considered in the strong coupling zero
approximation:

\begin{eqnarray}
\label{Eqn22} E_1 (0) \simeq  E^{(0)}_1 [U_{qs}; V_{qs};\varphi(\vec{r})] = <\Phi^{(0)}_1|\hat H |\Phi^{(0)}_1>,
\end{eqnarray}

where the average is calculated with the full hamiltonian (\ref{Eqn18}) and the functions $U_{qs};
V_{qs};\varphi(\vec{r})$ are to be found as the solutions of variational equations

\begin{eqnarray}
\label{Eqn23} \frac{\partial E^{(0)}_1 (U_{qs}; V_{qs};\varphi(\vec{r}))}{\partial U_{qs}} = \frac{\partial
E^{(0)}_1 }{\partial V_{qs}} =  \frac{\partial E^{(0)}_1 }{\partial \varphi(\vec{r})} = 0
\end{eqnarray}

under some additional conditions (\ref{Eqn21} - \ref{Eqn21a}).

It is quite natural, that the ground state energy is calculated  in the framework of the considered approximation
as follows:

\begin{eqnarray}
\label{Eqn24} E_0 \simeq  E^{(0)}_0 = <\Phi_0|\hat H |\Phi_0>.
\end{eqnarray}

Then the energy of the quasi-particle representing the "physical" electron (positron) is calculated by the
formula

\begin{eqnarray}
\label{Eqn24a} E(\vec P= 0) = E^{(0)}_1 -  E^{(0)}_0.
\end{eqnarray}

Further discussion is needed in connection with the application of the variational principle
(\ref{Eqn22}-\ref{Eqn23}) for estimating the energy of the excited state, because usually the variational
principle is used for estimating the ground state energy only. As far as we know it was first applied in
\cite{Caswell}, \cite{Yukalov} for nonperturbative calculation of the excited states of the anharmonic oscillator
with an arbitrary value of anharmonicity . This approach was called the "principle of the minimal sensitivity".
It was shown in our paper  \cite{OM82}, that the application of the variational principle to the excited states
is actually the consequence of the fact that the set of eigenvalues for the full Hamiltonian doesn't depend on
the choice of the representation for eigenfunctions. As a result, the operator method (OM) for solving
Schr\"odinger equation was developed as the regular procedure for calculating further corrections to the
zero-order approximation. Later OM was applied to a number of real physical systems and proved to be very
effective when calculating the energy spectrum in the wide range of the Hamiltonian parameters and quantum
numbers (\cite{OM} and the cited references). These results can serve as the background for the application of
the variational estimation to the excited state energy in QED.

The average value in Eq. (\ref{Eqn22}) is calculated neglecting the classical components of the vector field (its
quantum fluctuations should be taken into account in the further approximations)

\begin{eqnarray}
\label{Eqn25} <\Phi^{(0)}_1 | \hat \psi^* (\vec{r}) [ \vec \alpha \hat{ \vec{ A} } (\vec{r})] \hat \psi
(\vec{r})|\Phi^{(0)}_1> = 0.
\end{eqnarray}

It should be noted that the possibility of constructing self-consistently the renormalized QED at the non-zero
vacuum value of the scalar field operator was considered before  \cite{Fradkin} but the solution itself of the
corresponding equations was not  discussed.

It is also essential that the normal ordering of the operators  wasn't introduced in the hamiltonian
(\ref{Eqn18}). Therefore, its average value includes the infinite values corresponding to the vacuum energies of
electron-positron field for both the ground and the excited states. However, if the normalized conditions
(\ref{Eqn21}) are fulfilled these values are compensated in expression (\ref{Eqn24a}) for the quasi-particle
energy. Then the functional for determining the zero approximation for the energy of the one-particle excitation
is defined as follows:

\begin{eqnarray}
\label{Eqn25a} E(\vec P=0) = \int d \vec r \int \frac{d \vec q}{(2\pi)^{3/2}} \int \frac{d \vec
{q}'}{(2\pi)^{3/2}} \sum_{s,s'} \sum_{\mu,\nu} \{ U^*_{q's'} u^*_{\vec{q}' s' \mu} [(\vec \alpha \vec q + \beta
m_0)_{\mu \nu} + \nonumber\\
e_0 \varphi (\vec r) \delta_{\mu \nu}]
U_{q s} u_{\vec{q} s \nu} - \nonumber\\
V_{q's'} v^*_{\vec{q}' s' \mu} [(\vec \alpha \vec q + \beta m_0)_{\mu \nu} + e_0 \varphi (\vec r) \delta_{\mu
\nu}] V^*_{qs} v_{\vec{q} s \nu}\} e^{i (\vec q - \vec{q}') \vec r} - \nonumber\\
\frac{1}{2} \int d \vec r
[\vec{\nabla} \varphi (\vec r)]^2.
\end{eqnarray}

In order to vary the introduced functional let us define the spinor wave functions (not operators) which
describe the coordinate representation for the electron and positron wave packets in the state vector  $
|\Phi^{(0)}_1>$:

\begin{eqnarray}
\label{Eqn26} \Psi_{\nu} (\vec r) = \int \frac{d \vec q}{(2\pi)^{3/2}} \sum_{s} U_{q s} u_{\vec{q} s \nu}
e^{i \vec q \vec r}; \nonumber\\
\Psi^c_{\nu} (\vec r) = \int \frac{d \vec q}{(2\pi)^{3/2}} \sum_{s} V^*_{q s} v_{\vec{q} s \nu} e^{i \vec q \vec
r}.
\end{eqnarray}

In particular, if the trial state vector is chosen in one of the forms (\ref{Eqn21b}) of the standard
perturbation theory, the wave functions  (\ref{Eqn26}) coincide with the plane wave solutions of the free Dirac
equation. For a general case the variation of the functional (\ref{Eqn25a}) by the scalar field leads to

\begin{eqnarray}
\label{Eqn27} E(0) = \int d \vec r \{ \Psi^+ (\vec r) [(-i\vec \alpha \vec \nabla + \beta m_0) +
\frac{1}{2}e_0 \varphi (\vec r) ] \Psi (\vec r) - \nonumber\\
\Psi^{+c} (\vec r) [(-i\vec \alpha \vec \nabla + \beta m_0) +
\frac{1}{2}e_0 \varphi (\vec r) ] \Psi^c (\vec r); \nonumber\\
\int {d \vec{r}} [\Psi^{+} (\vec r) \Psi (\vec r) + \Psi^{+c} (\vec r') \Psi^{c} (\vec r')] = 1;
\end{eqnarray}

\begin{eqnarray}
\label{Eqn27b} \varphi (\vec r) = \frac{e_0}{4 \pi} \int \frac {d \vec{r}'} {|\vec r - \vec{r}'|} [\Psi^{+} (\vec
r') \Psi (\vec r') - \Psi^{+c} (\vec r') \Psi^{c} (\vec r')];
\end{eqnarray}
\begin{eqnarray}
\label{Eqn27c} \int {d \vec{r}} [\Psi^{+} (\vec r) \Psi (\vec r) + \Psi^{+c} (\vec r') \Psi^{c} (\vec r')] = 1.
\end{eqnarray}

The main condition for the existence of the considered nonperturbative excitation in QED is defined by the
extremum of the functional (\ref{Eqn27}) corresponding to a non-zero classical field. The structure of this
functional shows that such solutions of the variational equations could appear only if the trial state vector
simultaneously included the superposition of the electron and positron wave packets.

Equation (\ref{Eqn27}) and the Fourier representation  (\ref{Eqn20}) for the trial vector clearly indicate that
the assumption concerning the localization of the functions $\Psi (\vec r)$ near some point doesn't contradict
the translational symmetry of the system because this point  can itself be situated at any point of the full
space with equal probability. The general analysis of the correlation between the local violation of the symmetry
and the conservation of accurate integral of motion for the arbitrary quantum system was first discussed in
detail by Bogoluibov in his widely known paper "On quasi-averages" \cite{quasi}. A similar analysis of the
problem in question will be given in $\S$ 4.

Varying the functional (\ref{Eqn27}) by the wave functions $\Psi (\vec r)$ è $\Psi^c (\vec r)$ taking into
account their normalization leads to the equivalent Dirac equations  describing the electron (positron) motion in
the field of  potential $\varphi (\vec r)$:

\begin{eqnarray}
\label{Eqn27a} \{(-i\vec \alpha \vec \nabla + \beta m_0) +
e_0 \varphi (\vec r) \} \Psi (\vec r) = 0;\nonumber\\
\{(-i\vec \alpha \vec \nabla + \beta m_0) + e_0 \varphi (\vec r) \} \Psi^{c} (\vec r) = 0.
\end{eqnarray}

But it is important that in spite of the normalized condition (\ref{Eqn27c}) for the total state vector
(\ref{Eqn26}) each of its components could be normalized differently

\begin{eqnarray}
\label{Eqn27d} \int {d \vec{r}} \Psi^{+} (\vec r) \Psi (\vec r) = \frac{1}{1 + C}; \nonumber\\
\int {d \vec{r}}\Psi^{+c} (\vec r') \Psi^{c} (\vec r') = \frac{C}{1 + C}.
\end{eqnarray}

The constant  $C$ is arbitrary up to now. It defines the ratio of two charge states in the considered wave
packet. As a result the self-consistent potential  $\varphi (\vec r)$ of the scalar field depends on  $C$ because
of the equation (\ref{Eqn27b}).

We should discuss the procedure of separating variables in more detail, because of the non-linearity of the
obtained system of  equations for the wave functions and the self-consistent potential. Since the considered
physical system has no preferred vectors if  $\vec P = 0$, it is natural to regard the self-consistent potential
as spherically symmetrical. Then the variable separation for the Dirac equation is realized on the basis of the
well known spherical spinors \cite{Akhiezer}:

\begin{eqnarray}
\label{Eqn28}
\Psi_{jlM} = \left( \begin{array}{c}
g(r) \Omega_{jlM}\\
i f(r) \Omega_{jl'M}
\end{array} \right).
\end{eqnarray}

Here $\Omega_{jlM}$ are the spherical spinors \cite{Akhiezer} describing the spin and angular dependence of the
one-particle excitation wave functions; $j, M$ are the total excitation momentum and its projection respectively,
the orbital momentum eigenvalues are connected by the correlation $ l + l' = 2j$. It is natural to consider the
state with the minimal energy as the most symmetrical one, corresponding to the values  $j=1/2; M = \pm 1/2; l =
0; l' = 1$. This choice corresponds to the condition according to which in the non-relativistic limit the "large"
component of the spinor $\Psi$ $\sim g$ corresponds to the EPF electronic zone. Then the unknown radial functions
$f, g$ satisfy the following system of the equations:

\begin{eqnarray}
\label{Eqn29} \frac{d (rg)}{dr} - \frac{1}{r}(rg) - (E + m_0 - e_0 \varphi(r)) (rf) = 0;
\nonumber\\
\frac{d (rf)}{dr} + \frac{1}{r}(rf) + (E - m_0 - e_0 \varphi(r)) (rg) = 0.
\end{eqnarray}

The states with various projections of the total momentum should be equally populated in order to be consistent
with the assumption of the potential spherical symmetry with the equation (\ref{Eqn27b}). So, the total wave
function of the "electronic" component of the EPF quasi-particle excitation is chosen in the following form:

\begin{eqnarray}
\label{Eqn28a} \Psi = \frac{1}{\sqrt{2}} [\Psi_{1/2,0,1/2} + \Psi_{1/2,0,-1/2}]=
\left(
\begin{array}{c}
g(r) \chi^+_0\\
i f(r) \chi^+_1
\end{array} \right);
\nonumber\\
\chi^+_l =\frac{1}{\sqrt{2}} [\Omega_{1/2,l,1/2} + \Omega_{1/2,l,-1/2}]; \quad l = 0;1.
\end{eqnarray}

In its turn, the wave function  $\Psi^c$ is defined on the basis of the following bispinor:

\begin{eqnarray}
\label{Eqn30} \Psi^c_{jlM} = \left( \begin{array}{c}
- i f_1(r) \Omega_{jlM}\\
g_1(r) \Omega_{jl'M}
\end{array} \right).
\end{eqnarray}

The radial wave functions  $f_1, g_1$ in this case satisfy the following system of equations

\begin{eqnarray}
\label{Eqn31} \frac{d (rg_1)}{dr} + \frac{1}{r}(rg_1) - (E + m_0 + e_0 \varphi(r)) (rf_1) = 0;
\nonumber\\
\frac{d (rf_1)}{dr} - \frac{1}{r}(rf_1) + (E - m_0 + e_0 \varphi(r)) (r g_1) = 0.
\end{eqnarray}

These equations correspond to  the positronic branch of the EPF spectrum with the "large" component  $\sim g_1$
in the non-relativistic limit.

It is important to note that the functions  $\Psi$ and $\Psi^c$ satisfy the equations (\ref{Eqn27}) with the same
energy E(0). It imposes an additional condition of the orthogonality on them:

\begin{eqnarray}
\label{Eqn31a} <\Psi^c|\Psi>  = 0.
\end{eqnarray}

Taking into account this condition and also the requirement that the states with  different values of M should
be equally populated one finds the "positronic" wave function

\begin{eqnarray}
\label{Eqn31b} \Psi^c = \frac{1}{\sqrt{2}} [\Psi^c_{1/2,0,1/2} - \Psi^c_{1/2,0,-1/2}]= \left(
\begin{array}{c}
-i f_1(r) \chi^-_0\\
g_1(r) \chi^-_1
\end{array} \right );
\nonumber\\
\chi^-_l =\frac{1}{\sqrt{2}} [\Omega_{1/2,l,1/2} - \Omega_{1/2,l,-1/2}]; \quad l = 0;1.
\end{eqnarray}

The equation for the self-consistent potential follows from the definition of $\varphi (r)$ in  formula
 (\ref{Eqn27a})taking into account the normalization of the spherical spinors  \cite{Akhiezer}:

\begin{eqnarray}
\label{Eqn32} \frac{d^2 \varphi}{d r^2} + \frac{2}{r}\frac{d \varphi}{d r} = - \frac{e_0}{4\pi}[f^2 + g^2 -
f_1^2 - g_1^2].
\end{eqnarray}

The boundary condition for the potential is equivalent to the normalization condition (\ref{Eqn21a}) and defines
the charge of the "physical" electron (positron) e

\begin{eqnarray}
\label{Eqn33}
\varphi (r)|_{r \rightarrow \infty} = \frac{e}{4 \pi r} = \nonumber\\
\frac{e_0}{4\pi r}\int_{0}^{\infty} r^2_1 dr_1[f^2(r_1) + g^2(r_1) - f_1^2(r_1) - g_1^2(r_1)].
\end{eqnarray}

It is important to stress that the form of the functions given above is defined practically uniquely by the
imposed conditions. At the same time the obtained equations are consistent with the symmetries defined by the
physical properties of the system. The first symmetry is quite evident and relates to the fact that the
excitation energy doesn't depend on the choice of the quantization axis of the total angular momentum.

Moreover, these equations satisfy the condition of the charge symmetry \cite{Akhiezer}. Indeed, by direct
substitution, one can check  that one more pair of bispinors leads to the equations completely coinciding with
(\ref{Eqn29}),(\ref{Eqn31})

\begin{eqnarray}
\label{Eqn33a}\tilde{ \Psi}_{jlM} = \left( \begin{array}{c}
i g_1(r) \Omega_{jl'M}\\
- f_1(r) \Omega_{jlM}
\end{array} \right);
\end{eqnarray}

\begin{eqnarray}
\label{Eqn33b}\tilde{ \Psi}^c_{jlM}= \left( \begin{array}{c}
- f(r) \Omega_{jl'M}\\
- i g(r) \Omega_{jlM}
\end{array} \right).
\end{eqnarray}

It means that these bispinors allow one to find another pair of the wave functions which are orthogonal to each
other and to the functions  (\ref{Eqn28a}),(\ref{Eqn31b}) but include the same set of the radial functions

\begin{eqnarray}
\label{Eqn33ñ}
\tilde{\Psi} = \left(
\begin{array}{c}
i g_1(r) \chi^-_1\\
-f_1(r) \chi^-_0
\end{array} \right);\quad
\tilde{\Psi}^c =\left(
\begin{array}{c}
- f(r) \chi^+_1\\
-i g(r) \chi^+_0
\end{array} \right).
\end{eqnarray}

These functions differ from the set (\ref{Eqn28a}),(\ref{Eqn31b}) in that they have a different sign of the
observed charge of the quasi-particle due to the boundary condition  (\ref{Eqn33}).

Let us now proceed to the solution of the variational equations. It follows from the  qualitative analysis that
the important property of the trial state vector freedom is the relative contribution of the electronic and
positronic components of the wave function. Therefore let us introduce the variational parameter C in the
following way:

\begin{eqnarray}
\label{Eqn34}
\int_{0}^{\infty} r^2 dr[f^2(r) + g^2(r)] = \frac{1}{1 + C}; \nonumber\\
\int_{0}^{\infty} r^2 dr[f_1^2(r) + g_1^2(r)] = \frac{C}{1 + C}.
\end{eqnarray}

The dimensionless variables and new functions can be introduced

\begin{eqnarray}
\label{Eqn35} x = r m_0; \quad E = \epsilon m_0; \quad  e_0 \varphi(r) = m_0\phi(x);
\quad \frac{e^2_0}{4 \pi} = \alpha_0;\nonumber\\
u(x) \sqrt{m_0} = r g(r);\quad v(x)\sqrt{m_0} = r f(r);\quad u_1(x)\sqrt{m_0} = r g_1(r);\quad v_1(x)\sqrt{m_0}
= r f_1(r).
\end{eqnarray}

As a result the system of equations for describing the radial wave functions of the EPF one-particle excitation
and the self-consistent potential of the vacuum polarization can be obtained.

\begin{eqnarray}
\label{Eqn36} \frac{d u}{dx} - \frac{1}{x}u - (\epsilon + 1 - \phi(x)) v  = 0;
\nonumber\\
\frac{d v}{dx} + \frac{1}{x}v + (\epsilon - 1 - \phi(x)) u = 0;
\nonumber\\
\frac{d u_1}{dx} + \frac{1}{x}u_1 - (\epsilon + 1 + \phi(x)) v_1 = 0;
\nonumber\\
\frac{d v_1}{dx} - \frac{1}{x}v_1 + (\epsilon - 1 + \phi(x)) u_1 = 0;
\nonumber\\
\phi(x) = \alpha_0 [ \int_{x}^{\infty} dy \frac{\rho(y)}{y} +
\frac{1}{x} \int_{0}^{x} dy \rho(y)];\nonumber\\
\rho(x) = [u^2(x) + v^2(x) - u^2_1(x) - v^2_1(x)].
\end{eqnarray}

\section{Numerical solution of the self-consistent equations}

The mathematical structure of equations (\ref{Eqn36}) is analogous to that of the self-consistent equations for
localized state of "polaron" in the strong coupling limit  \cite{Bogol},\cite{Tyablikov}. Therefore the same
approach can be used for the numerical solution of these nonlinear equations. It has been developed and applied
\cite{Krylov} for the "polaron" problem on the basis of the continuous analog of Newton's method (CANM)
\cite{NAMN} .

Let us take into account that the system of equations (\ref{Eqn36}) can be simplified because the pairs of the
functions $u,v$ and $u_1,v_1$ are satisfied by the same equations and differ only by the normalized condition.
Therefore they can be represented by means of one pair of functions if special notations are used:

\begin{eqnarray}
\label{36d} u = \sqrt{\frac{1}{1 + C}} u_0; \ v = \sqrt{\frac{1}{1 + C}} v_0 ;
\nonumber\\
u_1 = \sqrt{\frac{C}{1 + C}} v_0; \ v_1 = \sqrt{\frac{C}{1 + C}} u_0 ;
\nonumber\\
\int_{0}^{\infty}  dx [u_0^2(x) + v_0^2(x)] = 1; \ \rho_0(x) = u_0^2(x) + v_0^2(x);
\nonumber\\
\frac{d u_0}{dx} - \frac{1}{x}u_0 - ( 1 - \phi(x)) v_0  = 0;
\nonumber\\
\frac{d v_0}{dx} + \frac{1}{x}v_0 - ( 1 + \phi(x)) u_0 = 0; \nonumber\\
\phi(x) = \alpha_0 \frac{1 - C}{1 + C}\phi_0(x); \ \phi_0(x) = [ \int_{x}^{\infty} dy \frac{\rho_0(y)}{y} +
\frac{1}{x} \int_{0}^{x} dy \rho_0(y)].
\end{eqnarray}

The energy of the system (\ref{Eqn27}) can also be calculated by these functions

\begin{eqnarray}
\label{37} E(0) = m_0 \frac{1 - C}{1 + C}[ T + \frac{1}{2}\alpha_0 \frac{1 - C}{1 + C} \Pi]; \nonumber\\
T = \int_{0}^{\infty} dx [ (u_0' v_0 - v_0' u_0) - 2 \frac{u_0 v_0}{x} + (u_0^2 - v_0^2)];
\nonumber\\
\Pi = \int_{0}^{\infty} dx \phi_0 (u_0^2 + v_0^2).
\end{eqnarray}
\noindent and equations (\ref{36d}) can be obtained as the result of the variation of the functional  (\ref{37}).

The required solutions are to be normalized and this condition defines the asymptotic behavior of the functions
near the integration interval boundaries:

\begin{eqnarray}
\label{37a} u_0 \approx A x[ 1 + \frac{1 - \phi^2(0)}{6}x^2] , \quad  v_0 \approx A \frac{1 - \phi(0)}{3} x^2,
\quad x \rightarrow 0;
\nonumber\\
u_0 \approx  A_1 \exp(-x), v_0 \approx  - A_1 \exp(-x)\quad x \rightarrow \infty.
\end{eqnarray}

The quantity

\begin{eqnarray}
\label{38}  a = \alpha_0 \frac{1 - C}{1 + C},
\end{eqnarray}
\noindent is a free parameter  and it should be chosen from the condition of the existence of non-zero and
normalized solution of  equations (\ref{36d}).

In order to use the method developed in ref.\cite{Krylov} especially for nonlinear eigenvalue problems let us
consider instead of  (\ref{36d}) the following system of equations:

\begin{eqnarray}
\label{36a} \frac{d u_k}{dx} - \frac{1}{x}u_k - ( k^2 - \phi(x)) v_k  = 0;
\nonumber\\
\frac{d v_k}{dx} + \frac{1}{x}v_k - ( k^2 + \phi(x)) u_0 = 0; \nonumber\\
\int_{0}^{\infty}  dx [u_k^2(x) + v_k^2(x)] = 1; \ \rho_k(x) = u_k^2(x) + v_k^2(x);
\nonumber\\
\phi(x) = a \phi_k(x); \ \phi_k(x) = [ \int_{x}^{\infty} dy \frac{\rho_k(y)}{y} + \frac{1}{x} \int_{0}^{x} dy
\rho_k(y)].
\end{eqnarray}

Here the eigenvalue $k^2(a)$ is introduced when the parameter  $a$ is fixed. In this case the normalization and
the boundary conditions

\begin{eqnarray}
\label{37b} u_k \approx A x[1 + \frac{k^2 - \phi_k^2(0)}{6}x^2] , \quad  v_0 \approx A \frac{k^2 - \phi(0)}{3}
x^2, \quad x \rightarrow 0;
\nonumber\\
u_k \approx   k A_1 \exp(-k x), v_k \approx  - A_1 \exp(-k x)\quad x \rightarrow \infty
\end{eqnarray}
\noindent allow one to calculate $k^2(a)$ by means of the numerical method from ref. \cite{Krylov}. Then one
should change the parameter  $a$ in order to find the value $a_0$ which satisfies the condition

\begin{eqnarray}
\label{37c} k^2(a_0) = 1.
\end{eqnarray}

Let us consider briefly the iteration procedure for calculating the functions $u_k(x),v_k(x)$ and the value  $k$.
If $(s)$ iterations are made, the desired values are defined by some approximate expression

\begin{eqnarray}
\label{00}  u^{(s)}_k(x), \ v^{(s)}_k(x) \ k^{(s)}; \nonumber\\
\int_0^{\infty}\rho^{(s)}_k(x) dx = 1, \ \rho^{(s)}_k(x) = [|u^{(s)}_k(x)|^2 + |v^{(s)}_k(x)|^2], \nonumber\\
\phi^{(s)}_k(x) = a[ \int_{x}^{\infty} dy \frac{\rho^{(s)}_k(y)}{y} + \frac{1}{x} \int_{0}^{x} dy
\rho^{(s)}_k(y)].
\end{eqnarray}

Then the following recurrent equations describe the wave functions and eigenvalues for the subsequent iterations:

\begin{eqnarray}
\label{01}  u^{(s+1)}_k(x) = A^{(s)}\{u^{(s)}_k(x) + \tau [ \psi^{(s)}(x) + \mu^{(s)} \psi^{(s)}_{\mu}(x)]\},
\nonumber\\
v^{(s+1)}_k(x) = A^{(s)}\{v^{(s)}_k(x) + \tau [ \psi^{(s)}_{1}(x) + \mu^{(s)} \psi^{(s)}_{1\mu}(x)]\}, \nonumber\\
k^{(s+1)} =  k^{(s)} + \mu^{(s)},
\end{eqnarray}
\noindent and the exact solutions are calculated as the limits of sequences

\begin{eqnarray}
\label{01a}  u_k(x) = \lim_{s \rightarrow \infty} u^{(s)}_k(x), \ v_k(x) = \lim_{s \rightarrow \infty}
v^{(s)}_k(x), \ k = \lim_{s \rightarrow \infty} k^{(s)}.
\end{eqnarray}

Here $\tau < 1$ is an arbitrary parameter determining the step of discretization for the evolutional equation
corresponding to CANM \cite{NAMN} for the considered problem (\ref{36d}). The corrections
$\psi^{(s)}(x),\psi^{(s)}_{1}(x),\psi^{(s)}_{\mu}(x),\psi^{(s)}_{1\mu}(x)$ are supposed to be small if the zeroth
approximation is good. They can be calculated as the solutions of the nonhomogeneous equations resulting from the
initial problem (\ref{36d}) \cite{Krylov}
\begin{eqnarray}
\label{02} \frac{d \psi^{(s)}}{dx} - \frac{1}{x}\psi^{(s)} - ( k^{(s)2} - \phi^{(s)}_k)\psi^{(s)}_{1}  = -
[\frac{d u^{(s)}_k}{dx} - \frac{1}{x}u^{(s)}_k - ( k^{(s)2} - \phi^{(s)}_k)v^{(s)}_0 ];
\nonumber\\
\frac{d \psi^{(s)}_1}{dx} + \frac{1}{x}\psi^{(s)}_{1} - (  k^{(s)2} + \phi^{(s)}_k)\psi^{(s)}  = - [\frac{d
v^{(s)}_k}{dx} + \frac{1}{x}v^{(s)}_k - ( k^{(s)2} + \phi^{(s)}_0)u^{(s)}_0 ];
\nonumber\\
\frac{d \psi^{(s)}_{\mu}}{dx} - \frac{1}{x}\psi^{(s)}_{\mu} - ( k^{(s)2} - \phi^{(s)}_k)\psi^{(s)}_{1\mu} = 2
k^{(s)}\phi^{(s)}_k v^{(s)}_k;
\nonumber\\
\frac{d \psi^{(s)}_{1\mu}}{dx} + \frac{1}{x}\psi^{(s)}_{1\mu} - ( k^{(s)2} + \phi^{(s)}_k)\psi^{(s)}_{\mu}  = 2
k^{(s)}\phi^{(s)}_k u^{(s)}_k.
\end{eqnarray}

The system of differential equations should be supplemented by the boundary conditions resulting from (\ref{37a})

\begin{eqnarray}
\label{02a} \psi^{(s)}_{1}(x) - \frac{k^{(s)2} + \phi^{(s)}_k(0)}{3} x \psi^{(s)}(x) = - [ v^{(s)}(x) -
\frac{k^{(s)2} + \phi^{(s)}_k(0)}{3} x u^{(s)}(x)], \ x \rightarrow 0,
\nonumber\\
k^{(s)}\psi^{(s)}_{1}(x) +  \psi^{(s)}(x) = - [ k^{(s)} v^{(s)}(x) + u^{(s)}(x)], \ x \rightarrow \infty, \nonumber\\
\psi^{(s)}_{1\mu}(x) - \frac{k^{(s)2} + \phi^{(s)}_k(0)}{3} x \psi^{(s)}_{\mu}(x) = 2 x \frac{k^{(s)}}{3}
u^{(s)}(x), \ x \rightarrow 0,
\nonumber\\
k^{(s)}\psi^{(s)}_{1\mu}(x) +  \psi^{(s)}_{\mu}(x) = - v^{(s)}(x), \ x \rightarrow \infty.
\end{eqnarray}

The correction $\mu^{(s)}$ for the eigenvalue $k^{(s)}$ and the normalization amplitude $A^{(s)}$ are calculated
on the basis of the normalization condition for the wave function. The first order correction for the
normalization integral tends to zero by means of the available choice of the value $\mu^{(s)}$ :

\begin{eqnarray}
\label{03} \mu^{(s)} = - \frac{I_{s}}{I_{\mu s}};
\nonumber\\
I_{s} = \int_0^{\infty}[ u^{(s)}_0 \psi^{(s)} + v^{(s)}_0\psi^{(s)}_{1}] dx;
\nonumber\\
I_{\mu s} = \int_0^{\infty}[ u^{(s)}_0 \psi^{(s)}_{\mu} + v^{(s)}_0\psi^{(s)}_{1\mu}] dx.
\end{eqnarray}
\noindent The coefficient $A^{(s)}$ allows one to normalize functions  (\ref{01}) for the subsequent iteration
with  the second order accuracy on the value $\mu^{(s)}$. The convergence of the considered iteration scheme was
proved in ref.\cite{Krylov} .

It is known \cite{log}, that the numerical solution of the differential equations (\ref{02}) within the interval
$[0,\infty]$ is more effective if the following logarithmic change of variables is used

\begin{eqnarray}
\label{04} x = e^{\theta}, \ \theta = \ln x, \ dx = e^{\theta} d\theta, \ -\infty < \theta < \infty,
\end{eqnarray}
\noindent It allows one to distribute the nodes for the finite-difference approximation of equations (\ref{02})
in a nonuniform way. As a result the step of integration is small in the range  $ x \leq 1$, where the desired
functions are most significant, and increases for large values of $x$ when the functions are small.

Zeroth approximation for the recurrent equations (\ref{02}) can be calculated by means of the variational
estimation for the functional (\ref{37}) using the following trial functions \cite{electron}

\begin{eqnarray}
\label{05} u_k^{(0)} = A x (1 + bx) e^{-bx}, \ v_k^{(0)} = B x^2 e^{-bx}, \nonumber\\
7 A^2 + 3 B^2 = 4 b^3.
\end{eqnarray}

Thus, for example, the recurrent sequence (\ref{01a}) converges quite quickly into the value $k (a) = 1.05$ for
the following initial values of the parameters

\begin{eqnarray}
\label{06} A =  \sqrt{2/7}, \ B = \sqrt{2/3},  \ k^{(0)} = 1, \ \tau = 0.5, \ b = 1, \ a = - 3.3.
\end{eqnarray}

It is important that the variation of the initial approximation for the functions and parameters affects the
speed of the iteration convergence but does not change the final result for $k (a)$. It corresponds to the only
global extremum for the functional (\ref{37}).

The numerical solutions $u_k(x) ,  v_k(x) , k (a)$ calculated with the fixed value $a$ were used further as the
initial approximation  for the iteration solution of the equations (\ref{02}) with the variation of the parameter
$a_1 = a + \Delta a$. This procedure was repeated until the condition (\ref{37c}) was satisfied with some given
accuracy. As a result $a_0$ was calculated up to 4 significant figures:

\begin{eqnarray}
\label{39}  a_0 \approx - 3.296,
\end{eqnarray}
\noindent and this value corresponds to the desired soliton-like solution with nonzero charge.

The numerical analysis of the equations (\ref{36d}) in  wide range of parameters has shown that the only solution
described above existed under the condition that the function $u_0$ should not have zero points for the ground
state. The value $a_0$ can also be expressed in terms of the integrals from the functional (\ref{37}) provided of
its minimum relative to the parameter $\delta = (1-C)/(1+C)$, which defines the contributions of the electron and
positron components to SLE :

\begin{eqnarray}
\label{40}  a_0 =  - \frac{T}{\Pi}; \quad E(0) = - \frac{m_0}{\alpha_0}\frac{T a }{2}.
\end{eqnarray}

The calculated value of the integral

\begin{eqnarray}
\label{40b}  T = 0.749.
\end{eqnarray}

Fig.\ref{Fig.1} shows the solutions $u_0, v_0$ of the Dirac equation and Fig.\ref{Fig.2} represents the
self-consistent potential $\phi_0$ corresponding to the value $a_0$. The characteristic size $\Delta r$ of the
spatial localization of the excitation is defined by the parameter

\begin{eqnarray}
\label{40a} \Delta r \approx m_0^{-1}.
\end{eqnarray}

\begin{figure}[h]\centering{
\includegraphics [scale=0.7]{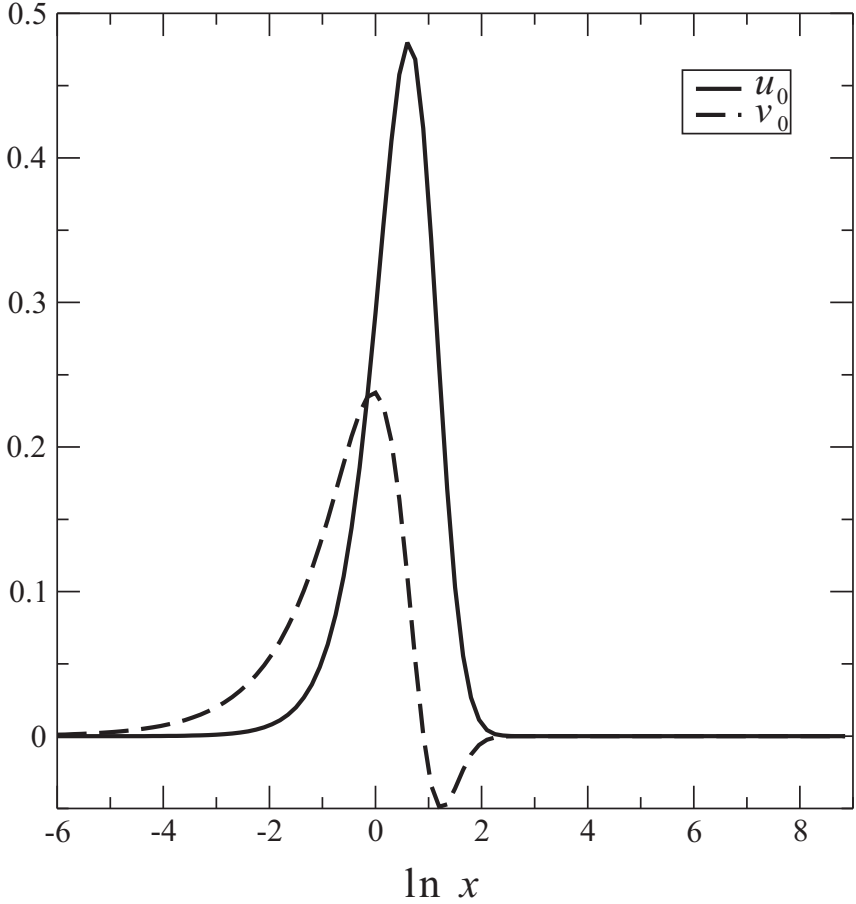}} \caption {Solutions of the Dirac equation for SLE} \label{Fig.1}
\end{figure}

\begin{figure}[h]\centering{
\includegraphics[scale=0.7]{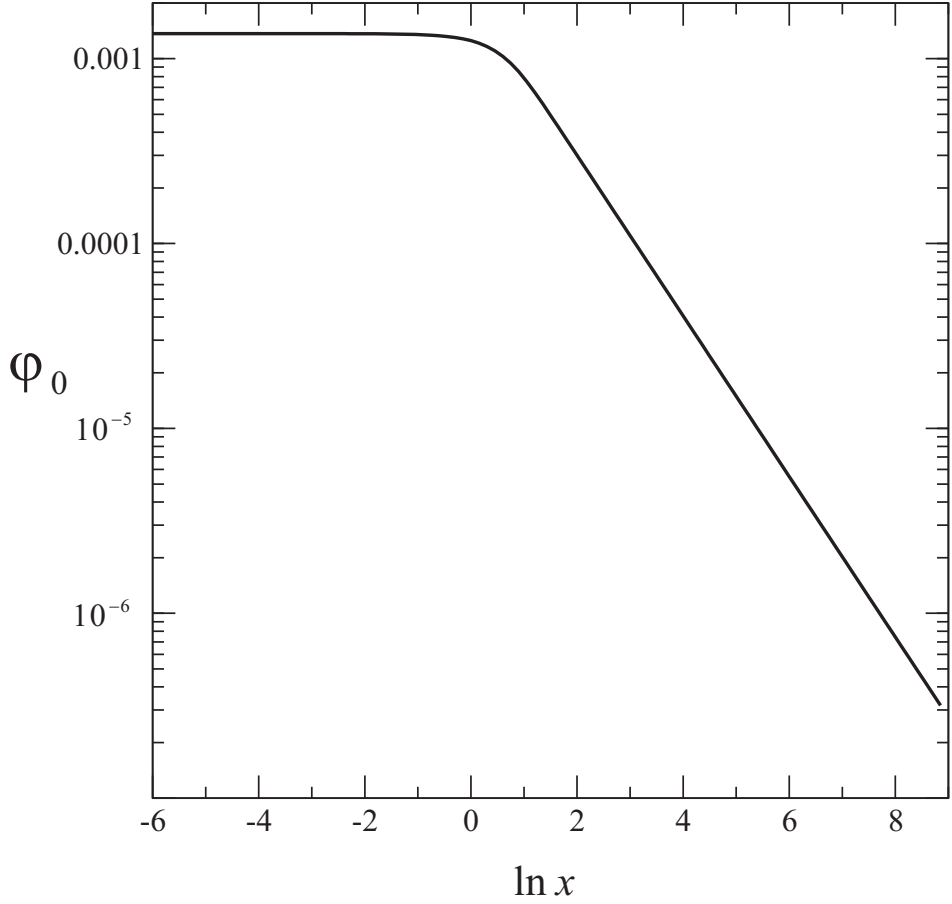}} \caption {Self-consistent potential of SLE} \label{Fig.2}
\end{figure}

The equations (\ref{Eqn33}) and (\ref{Eqn36}) lead to the following relation between the "bare" charge  $e_0$ and
the integral charge of SLE:

\begin{eqnarray}
\label{41} e_0 \frac{(1 - C)}{(1 + C)}  = e; \nonumber\\
e \ e_0 = 4 \pi a, \quad e = \frac{4 \pi a}{e_0} \approx - \frac{41.42}{e_0}.
\end{eqnarray}

It is of interest that the formula (\ref{41}) has the same structure and consequences that result from the Dirac
monopole theory\cite{monopol} despite of a different physical interpretation. In particular, if the calculated
self-localized one-particle excitation of the electron-positron field (EPF) could be considered as a "physical"
electron the only possible value of the parameter $a_0$ together with the ratio (\ref{41}) leads to the condition
of the observed charge quantization with the fixed value of the "bare" charge $e_0$. Indeed, on the analogy with
the "polaron" theory \cite{bipolaron},\cite{Solodovnikova1} n-particle excitation of EPF can be considered as the
superposition of n different SLE' that are at the distances exceeding the size of localization $\Delta r $. Then
the charge of such n-soliton excitation $q_n$ would be always multiple to the elementary charge of SLE $q_n = n
e$.

\section{EPF quasi-particle excitation with an arbitrary momentum}

In the previous section we have considered  the possibility for a resting quasi-particle with a non-trivial
self-consistent charge distribution, the finite energy $E(0)$ and a zero total momentum $\vec P = 0$ to exist in
the framework of a nonperturbative QED.

The obtained solution allows one to imagine the internal structure of the resting "physical" electron (positron)
as a strongly coupled state of charge distributions of the opposite sign, the large values of integral charges of
these distributions compensating each other almost completely and their heavy masses being "absorbed" by the
binding energy.

Actually the energy $\pm E(0)$ defines the boundaries of the renormalized electron and positron zones resulting
from the strong polarization of EPF when the excitation appears. But this excitation could be interpreted as the
"physical" electron (positron) if the sequence of the levels in every zone determined by the vector $\vec P \not=
0 $ (Fig.1) were described by the relativistic energy spectrum of real particles, that is

\begin{eqnarray}
\label{Eqn47} E(\vec P) = \sqrt{ P^2 + E^2 (0) } = \sqrt{ P^2 + m^2 }.
\end{eqnarray}

Only in this case the energy E(0) can really be used for  calculating the mass $m_0$ of the "bare" electron.

It is worth saying, that the problem of studying the dynamics of the self-localized excitation should be solved
for any system with a strong interaction between quantum fields in order to calculate its effective mass. For
example, a similar problem for Pekar "polaron"   \cite{Pekar} in the ionic crystal was considered in papers
\cite{PV},\cite{Bogol}, \cite{Tyablikov}, \cite{Feynmanpol}, \cite{Gross} and in many more recent works. It is
essential that because of the non-linear coupling between the particle and a self-consistent field the energy
dispersion $E(\vec P)$ for the quasi-particle proves to be very complicated. As the result, its dynamics in the
crystal is similar to the motion of the point  "physical" particle only at a small enough total momentum.

However, in the case of QED the problem is formulated in a fundamentally different way. There is currently no
doubt that the dynamics of the "physical" excitation should be described by the formula (\ref{Eqn47}) for any (!)
values of the momentum $\vec P$. It means that the considered nonperturbative approach for describing the
internal structure of the "physical" electron should lead to the energy dispersion law (\ref{Eqn47}) for the
entire range of the momentum $\vec P$.

The rigorous method of taking into account the translational symmetry in the strong coupling theory for the
"polaron" problem was elaborated in the works of Bogolubov \cite{Bogol} and Gross \cite{Gross}. Let's remember
that this method was based on the introduction of the collective variable  $\vec R$ conjugated to the total
momentum operator  $ \hat {\vec P}$, the canonical character of the transformation caused by three new variables
$R_i$ being provided by the same number of additional conditions imposed on the other variables of the system. In
the "polaron" problem the quantum field interacting with the particle contributes to the total momentum of the
system. It allows one to impose these conditions on the canonical field variables \cite{Bogol}, \cite{Gross} and
the concrete form of the variable transformation is based mainly on the permutation relations for the boson field
operators.

The considered problem has some specific features in comparison with the "polaron" problem. Firstly, the
formation of the one-particle wave packet is the multi-particle effect because this packet includes all initial
states of EPF as the fermionic field. Secondly, its self-localization is provided by the polarization potential
of the scalar field that doesn't contribute to the total momentum of the system. Therefore, we use a different
approach in order to select the collective coordinate  $\vec R$. Let us return to the configuration
representation in the Hamiltonian  (\ref{Eqn18}) , where QED is considered to be the totality of N $( N
\rightarrow \infty)$ point electrons interacting with the quantum EMF in the Coulomb gauge \cite{Heitler}:

\begin{eqnarray}
\label{10} \hat H = \sum_{a=1}^{N} \{  \vec{\alpha}_a [\hat { \vec{p}_a} + e_0 \hat{ \vec{ A} } (\vec{r}_a)] +
\beta_a m_0 + e_0 \hat \varphi (\vec{r}_a)\} - \nonumber\\  \frac{1}{2} \int d \vec{r}( \vec{\nabla}\hat \varphi
(\vec{r}))^2 + \sum_{\vec k \lambda} \omega (\vec k) \hat n_{\vec k \lambda};
\nonumber\\
\omega (\vec k) = k; \quad \hat n_{\vec k \lambda} = c^+_{\vec k \lambda}c_{\vec k \lambda}; \quad \lambda = 1,2;
\nonumber\\
\hat{ \vec{ A} } (\vec{r}) = \sum_{\vec k \lambda} \frac{1}{\sqrt{2 k \Omega}} \vec {e}^{(\lambda)} [ c_{\vec k
\lambda} e^{i \vec k \vec r} + c^+_{\vec k \lambda} e^{-i \vec k \vec r}]
\end{eqnarray}

Here $\Omega$ is the normalized volume; $c^+_{\vec k \lambda}(c_{\vec k \lambda})$ are the operators of the
creation (annihilation) of quanta of a transversal electromagnetic field, the quantum having the wave vector
$\vec k$, polarization  $\vec {e}^{(\lambda)}$ and energy $\omega (\vec k) = k$. The sign of the interaction
operators differs from the standard one because the parameter $e_0$ is introduced as a positive quantity.

As it was stated above the zero approximation of nonperturbative QED is defined only by a strong interaction of
electrons with the scalar field, where the interaction with the transversal field is to be taken into account in
the framework of the standard perturbation theory when the conservation of the total momentum is provided
automatically \cite{Akhiezer}. Therefore, while describing the quasi-particle excitation we should consider the
conservation of the total momentum only for the system of electrons. In the considered representation it is
defined by the sum of the momentum operators of individual particles

\begin{eqnarray}
\label{11} \hat {\vec P} = \sum_{a=1}^{N} \hat {\vec{p}_a};  \nonumber\\
\hat {\vec P} |  \Phi_1 (\vec P)> = \vec P |  \Phi_1 (\vec P)>.
\end{eqnarray}

It means that in the configuration space the variable $\vec R$ conjugated to the total momentum is simply a
coordinate of the center of mass and the desired transformation to new variables is as follows:

\begin{eqnarray}
\label{12}\vec r_a = \vec R + \vec{\rho}_a; \quad  \vec R = \frac{1}{N}\sum_{a=1}^{N}\vec r_a; \quad
\sum_{a=1}^{N} \vec{\rho}_a = 0; \nonumber\\
\hat {\vec{p}_a} = - i \vec {\nabla}_a = - \frac{i}{N}\vec {\nabla}_R + \hat {\vec{p}'_a}; \quad \hat {\vec P} =
-i\vec {\nabla}_R; \quad \hat {\vec{p}'_a} =- i \vec {\nabla}_{\rho_a} + \frac{i}{N}\sum_{b=1}^{N}\vec
{\nabla}_{\rho_b}; \quad \sum_{a=1}^{N} \hat {\vec{p}'_a} = 0.
\end{eqnarray}

The Hamiltonian (\ref{10}) with new variables is of the following form

\begin{eqnarray}
\label{13} \hat H = \sum_{a=1}^{N} \{  \vec{\alpha}_a [- \frac{i}{N}\vec {\nabla}_R + \hat {\vec{p}'_a}
 + e_0 \hat{ \vec{ A} } (\vec{\rho}_a + \vec R)] + \beta_a m_0 + e_0 \hat \varphi (\vec{\rho}_a + \vec R)\}
 - \nonumber\\
\frac{1}{2} \int d \vec{r}( \vec{\nabla}\hat \varphi (\vec{r}))^2 + \sum_{\vec k \lambda} \omega (\vec k) \hat
n_{\vec k \lambda}.
\end{eqnarray}

It should be noted that the matrix elements of an arbitrary operator in a new configuration representation are to
be calculated in accordance with the following norm (we introduce a special notation for this norm)

\begin{eqnarray}
\label{14} << \Phi_1 | \hat M | \Phi_2 >> = \int d \vec R \prod_{a} d \vec {\rho}_a \Phi_1^* (\vec R, \{\vec
{\rho}_a\})\hat M \Phi_2 (\vec R, \{\vec {\rho}_a\}).
\end{eqnarray}

Let us denote by $\hat H_0$ that part of the operator (\ref{13}) which doesn't depend on the transversal EMF and
describes the internal structure of "physical" particles. In fact, the operator  $\hat H_0$ doesn't depend  on
the coordinate $\vec R$ either, because of its commutativity with the operator of the total momentum of the
system of electrons. This also follows from the well known result  \cite{Heitler} that in the Coulomb gauge the
scalar potential could be excluded from the hamiltonian. As a result the operator $\hat H_0$ depends only on the
vector differences  $(\vec {r}_a - \vec {r}_b) = (\vec {\rho}_a - \vec {\rho}_b)$ and doesn't change with the
simultaneous translation of all the coordinates. As a consequence, the eigenfunctions of the hamiltonian $\hat
H_0$ depend on the coordinate $\vec R$ in the same way as for a free particle:

\begin{eqnarray}
\label{15}  \Phi (\vec R, \{\vec {\rho}_a\}) = \frac{1}{(2\pi)^{3/2}}e^{i\vec P\vec R} |\Phi_1 (\vec
P, \{\vec {\rho}_a\})>;\nonumber\\
\hat H_0 \rightarrow \hat H_0 (\vec P) = \sum_{a=1}^{N} \{  \vec{\alpha}_a [ \frac{1}{N}\vec {P} + \hat
{\vec{p}'_a} ] + \beta_a m_0 + e_0 \hat \varphi (\vec{\rho}_a )\} -  \frac{1}{2} \int d \vec{r}(
\vec{\nabla}\hat \varphi (\vec{r}))^2
\end{eqnarray}

Further calculations consist in returning to the field representation by the variables   $ \vec {\rho}_a$ in the
limit $( N \rightarrow \infty)$ and in using the approximate trial wave packet $|\Phi_1 (\vec P)>$ similar to
(\ref{Eqn20}) but with the coefficient functions depending on  $\vec P$ . Thus, in the framework of
non-perturbation QED the orthogonalized and normalized set of states for the EPF one-particle excitation is
defined as follows

\begin{eqnarray}
\label{16} |\Phi^{(0)}_1(\vec P) > \simeq \frac{1}{(2\pi)^{3/2}}e^{i\vec P\vec R} \int d \vec{q} \{
U_{\vec{q}s}(\vec P) a^+_{\vec{q} s} + V_{\vec{q}s}(\vec P) b^+_{\vec{q} s} \} | 0; 0;\varphi(\vec r)>;
\end{eqnarray}

with the norm (\ref{14}) and the coefficient functions  $U_{\vec{q}s}(\vec P);V_{\vec{q}s}(\vec P)$.

Using the coordinate representation for these functions

\begin{eqnarray}
\label{Eqn48} \Psi_{\nu} (\vec r, \vec P) = \int \frac{d \vec q}{(2\pi)^{3/2}} \sum_{s} U_{\vec q s}(\vec P)
u_{\vec{q} s \nu}
e^{i \vec q \vec r}; \nonumber\\
\Psi^c_{\nu} (\vec r,\vec P ) = \int \frac{d \vec q}{(2\pi)^{3/2}} \sum_{s} V_{\vec q s}(\vec P) v_{\vec{q} s
\nu} e^{i \vec q \vec r},
\end{eqnarray}
\noindent one can find the following functional for calculating the value $E(\vec P) = E_1^{(0)} - E_0^{(0)}$
corresponding to the energy of one-particle excitation with an arbitrary momentum

\begin{eqnarray}
\label{Eqn49} E(\vec P) = \int d \vec r \{ \Psi^+ (\vec r, \vec P) [(\vec \alpha \vec P -i\vec \alpha \vec \nabla
+ \beta m_0) + e_0\frac{1}{2} \varphi (\vec r, \vec P) ] \Psi (\vec r, \vec P) -
\nonumber\\
\Psi^{+c} (\vec r, \vec P) [(- \vec \alpha \vec P -i\vec \alpha \vec \nabla + \beta m_0) +
e_0\frac{1}{2} \varphi (\vec r, \vec P) ] \Psi^c (\vec r, \vec P); \nonumber\\
\varphi (\vec r, \vec P) = e_0 \int \frac {d \vec{r}'} {|\vec r - \vec{r}'|} [\Psi^{+} (\vec r', \vec P) \Psi
(\vec r', \vec P) - \Psi^{+c} (\vec r', \vec P) \Psi^{c} (\vec r', \vec P)];
\nonumber\\
\int {d \vec{r}} [\Psi^{+} (\vec r, \vec P) \Psi (\vec r, \vec P) + \Psi^{+c} (\vec r', \vec P) \Psi^{c} (\vec
r', \vec P)] = 1.
\end{eqnarray}

It should be noted that the mean value of the operator $\hat H_0 (\vec P)$ in the representation of the second
quantization was calculated by taking into account the expression

$$
\lim_{N \to \infty} \frac{1}{N} \sum_{\vec p} = 1,
$$
\noindent that follows from a well known formula for the density of states when EPF is quantized within the
normalized volume $\Omega$ \cite{Akhiezer}.

It was mentioned above, that there is quite a close analogy between the nonperturbative description of QED and
the theory of strong coupling in the "polaron" problem. Therefore, it is of interest to compare the obtained
functional (\ref{Eqn49})with the results of various methods of including the translational motion in the
"polaron" problem. The simplest one was used by Landau and Pekar \cite{PV} who introduced the Lagrange
multipliers in the form

\begin{eqnarray}
\label{Eqn49a} J(\vec P) = J( \vec P =0) + (\vec P \vec V),
\end{eqnarray}

with the functional $J( \vec P =0)$ referring to a resting "polaron" and the Lagrange multiplier  $V_i$ denoting
the components of the quasi-particle average velocity.

We can see that the obtained functional(\ref{Eqn49}) has the same form as  (\ref{Eqn49a})if the relativistic
velocity of the excitation is determined by the formula

\begin{eqnarray}
\label{Eqn49b} \vec V = \int d \vec r \{\Psi^+ (\vec r, \vec P) (\vec \alpha)  \Psi (\vec r, \vec P) + \Psi^{+c}
(\vec r, \vec P) (\vec \alpha)  \Psi^c (\vec r, \vec P)\} .
\end{eqnarray}

This corresponds to the well known interpretation of  Dirac matrixes.

Varying this functional by taking into account the normalized conditions leads to the following equations for the
wave functions

\begin{eqnarray}
\label{Eqn50} \{(\vec \alpha \vec P -i\vec \alpha \vec \nabla + \beta m_0) + e_0 \varphi (\vec r, \vec P) -
E(\vec P)\} \Psi (\vec r, \vec P) = 0;
\nonumber\\
\{(\vec \alpha \vec P + i\vec \alpha \vec \nabla - \beta m_0) - e_0 \varphi (\vec r, \vec P) - E(\vec P) \}
\Psi^c (\vec r, \vec P)= 0.
\end{eqnarray}

These equations show that  unlike the "polaron" problem \cite{Bogol}, the translational motion of the
quasi-particle determining the momentum $\vec P$ is related in our case to its internal movement described by the
coordinate $\vec r$  by means of spinor variables only. The physical reason for this separation of variables is
explained by the fact that in QED the self-localized state is formed by the scalar field, and its interaction
with the particle doesn't involve the momentum exchange. In order to find the analytical energy spectrum $E(\vec
P)$ the system of non-linear equations(\ref{Eqn50} should be diagonalized relative to the spinor variables. The
possibility of such diagonalization seems to be a non-trivial requirement for the nonperturbative QED under
consideration.

The solution of the equations  (\ref{Eqn50}) can be found on the basis of the states for which the dependence on
the vector $\vec q$ in the wave packet amplitudes  $U_{\vec q, s}, V_{\vec q, s}$ remains the same as it was in
the motionless "physical" electron. However, the relation between the spinor components of these functions can be
changed. But as the self-consistent scalar potential involves the summation over all spinor components it can be
assumed that the potential doesn't depend on the momentum for the class of states in question:

\begin{eqnarray}
\label{Eqn51} \varphi (\vec r, \vec P) = \varphi (r) |_{\vec P = 0}.
\end{eqnarray}

In the coordinate representation the transformation of the spinor components of the wave functions satisfying the
equations (\ref{Eqn50}) takes place because of the dependence on the momentum. It seems that there are some
general arguments based on the Hamiltonian symmetry that make the diagonalization of equations (\ref{Eqn50})
easier. However, we have desired the solution by sorting out various linear combinations of the wave functions
found in Section 3 for a resting electron. These functions correspond to the degenerated states in the case of
$\vec P = 0$ but are mixed for a moving electron. It is found that there is only one normalized linear
combination satisfying all the necessary conditions of self-consistency :

\begin{eqnarray}
\label{Eqn52} \Psi (\vec r, \vec P)  = L (\vec P) \Psi (\vec r) + K (\vec P) \tilde{\Psi^c} (\vec r) ;
\nonumber\\
\Psi^c(\vec r, \vec P)  = L_1(\vec P) \Psi^c (\vec r) +
K_1(\vec P) \tilde{\Psi}(\vec r); \nonumber\\
|L|^2 + |K|^2 = |L_1|^2 + |K_1|^2 = 1.
\end{eqnarray}

The condition  (\ref{Eqn51}) according to which the potential doesn't depend on the momentum for the excitation
at the energy $E(\vec P)$, is fulfilled if the coefficients are related as

\begin{eqnarray}
\label{Eqn52a} L_1 = - K; \qquad K_1 = L.
\end{eqnarray}

These relations are also consistent with  Equations (\ref{Eqn50}) for  wave functions.

This means that the "physical" electron moves in such a way that its states are transformed in the phase space of
the orthogonal wave functions (\ref{Eqn28a}),(\ref{Eqn31b}),(\ref{Eqn33b}),(\ref{Eqn33ñ})but the amplitudes of
its "internal" charge distributions are not changed. These results, however, are valid only for the case of
neglecting the interaction with the transverse electromagnetic field.

Substituting the superpositions  (\ref{Eqn52}) into equations (\ref{Eqn50}) we use the following relations:

\begin{eqnarray}
\label{Eqn53} (\vec \alpha \vec P) \Psi(\vec r) = \left( \begin{array}{c}
i f (r) (\vec \sigma \vec P)\Omega_{1/2, 1, M}\\
g(r) (\vec \sigma \vec P) \Omega_{1/2,0,M}
\end{array} \right); \quad
(\vec \alpha \vec P) \Psi^c(\vec r) = \left( \begin{array}{c}
g_1 (r) (\vec \sigma \vec P)\Omega_{1/2, 1, M}\\
-i f_1(r) (\vec \sigma \vec P) \Omega_{1/2,0,M}
\end{array} \right);\nonumber\\
(- i \vec \alpha \vec \nabla + \beta m_0 + e_0 \varphi) \Psi(\vec r) = E(0) \left( \begin{array}{c}
g (r) \Omega_{1/2, 0, M}\\
i f(r) \Omega_{1/2,1,M}
\end{array} \right); \nonumber\\
(-i \vec \alpha \vec \nabla + \beta m_0 + e_0 \varphi)  \Psi^c(\vec r) = - E(0) \left( \begin{array}{c}
-i f_1 (r) \Omega_{1/2, 0, M}\\
g_1(r) \Omega_{1/2,1,M}
\end{array} \right),
\end{eqnarray}

and similar formulas for the functions  $\tilde{\Psi}(\vec r); \tilde{\Psi^c}(\vec r)$;  $\sigma_i$ are the Pauli
matrixes.

For equations  (\ref{Eqn50}) to be fulfilled for any vector $\vec r$ it is necessary to set the coefficients of
spherical spinors equal to the same indexes  $l$. The corresponding radial functions are proved to be the same
under these conditions, and the following system of equations for the spinors $\chi_{0,1}^{\pm}$ is obtained

\begin{eqnarray}
\label{Eqn54} i L (\vec \sigma \vec P)\chi^+_1 + K ( E + E_0)\chi_1^+ = 0; \quad i K (\vec \sigma \vec P)\chi^+_0
+ L(
E - E_0)\chi_0^+ = 0;\nonumber\\
L (\vec \sigma \vec P)\chi^+_0 + i K ( E + E_0)\chi_0^+ = 0; \quad  K (\vec \sigma \vec P)\chi^+_1 + i L(
E - E_0)\chi_1^+ = 0;\nonumber\\
L_1 (\vec \sigma \vec P)\chi^-_1 - i K_1 ( E - E_0)\chi_1^- = 0; \quad  K_1 (\vec \sigma \vec P)\chi^-_0 - i L_1(
E + E_0)\chi_0^- = 0;\nonumber\\
i L_1 (\vec \sigma \vec P)\chi^-_1 -  K_1 ( E - E_0)\chi_1^- = 0; \quad i K_1 (\vec \sigma \vec P)\chi^-_0 -L_1(
E + E_0)\chi_0^- = 0.
\end{eqnarray}

Spin variables in Equations (\ref{Eqn54}) are also separated. In order to show this one can use, for example, the
relation between the coefficients resulting from the 4th equation in  (\ref{Eqn54}) in the first one of these
equations:

$$
\chi_1^+ = i \frac{K (\vec \sigma \vec P)\chi^+_1}{L( E - E_0)}.
$$

As aresult there exists a non-trivial solution of these equations for two branches of the energy spectrum

\begin{eqnarray}
\label{Eqn55} E_{e,p} = \pm \sqrt{E^2_0 + P^2},
\end{eqnarray}

referring to the electron and positron zones, respectively (Fig.1). The same expressions can be obtained for all
conjugated pairs of the equations in (\ref{Eqn54}). The coefficients in the wave functions (\ref{Eqn52}) can be
found taking into consideration the normalization condition:

\begin{eqnarray}
\label{Eqn56} L^e = K^e_1 = \frac{P}{\sqrt{P^2 + (E_e - E_0)^2}}; \quad K^e = - L^e_1 =
\frac{E_e - E_0}{\sqrt{P^2 + (E_e - E_0)^2}}; \nonumber\\
L^p = K^p_1 = \frac{P}{\sqrt{P^2 + (E_e + E_0)^2}}; \quad K^p = - L^p_1 = -\frac{E_e + E_0}{\sqrt{P^2 + (E_e  +
E_0)^2}}.
\end{eqnarray}

One can see that this set of coefficients coincides with the set of spinor components for solving the Dirac
equation for a free electron with the observed mass $ m = E_0$. Thus, the results of this section show that the
"internal" structure of the "physical" electron (positron) considered in this paper is consistent with the
experimental energy dispersion (\ref{Eqn55}) for a real free particle due to the relativistic invariance of the
Dirac equation.

For the interpretation of the wave packet (\ref{16}) as the state vector for a "physical" particle it is
essential that it is the eigenvector of the total momentum of EPF in the framework of the considered "quenched
QED". It means that SLE with different momenta form the set of orthogonal and normalized functions if the
condition (\ref{14}) is taken into account:

\begin{eqnarray}
\label{Eqn56a} << \Phi^{(0)}_1 (\vec P_1) |  \Phi^{(0)}_1 (\vec P)>> = \delta (\vec P - \vec P_1).
\end{eqnarray}

\section{Conclusions}

Thus, in the present paper new variational self-consistent equations for interacting electron-positron and scalar
electromagnetic fields have been derived in the framework of the "large-$\alpha$ QED". These equations differ
from the Schwinger-Dyson equations and describe the one-particle excitation of the system that doesn't contain
the quanta of the transversal electromagnetic field.

The soliton-like solution of the desired equations has been calculated numerically. It includes a non-zero
integral charge ("charge" soliton) and can be regarded as the spatially localized collective excitation of the
electron-positron field coupled with the localized scalar field. It is shown that this solution is unique if the
coupling constant $\alpha_0$ for the field interaction is fixed. This result leads to the quantization condition
for the "observed" charge of  $n$-soliton excitations of the system.

The dependence of the energy of the soliton-like excitation on its total momentum has been also studied. It is
shown that the corresponding law of the energy dispersion is consistent with Lorentz invariance of QED and the
effective mass of the "charge soliton" can be calculated.

It should be emphasized that the main aim of the paper is to prove the possibility of the existence of such
non-trivial solutions of the "large-$\alpha$ QED" equations that could not be derived on the basis of the
standard perturbation theory. However, the physical interpretation of the desired solution requires further
research.

\end{Large}


\begin{thebibliography}{}


\bibitem{large} J.R.Hiller and S.J.Brodsky, {\it Phys. Rev. D},
{\bf 59},(1998), 016006.


\bibitem{Kleinert} Proceedings: {\it Fluctuating Paths and Fields}, (Singapore,
World Scientific, 2001).

\bibitem{QCD} S.P.Klevansky, {\it Rev. of Modern Phys},
{\bf 64},(1992), 649.




\bibitem{SD} J.Schwinger, {\it Phys. Rev. },
{\bf 125},(1962),397; R.Fukuda and R.Kugo, {\it Nucl. Phys. }, {\bf B117},(1976),250;P.I.Fomin, V.P.Gusynin,
V.A.Miransky and Y.A.Sytenko, {\it Riv. Nuovo Chim. }, {\bf 6},(1983),1.

\bibitem{SD1} Y.Kikichi and Y.Ng, {\it Phys.Rev. D }, {\bf 38},(1988),3578;  V.P.Gusynin,
V.A.Miransky and I.A.Shovkovy, {\it Phys.Rev. D }, {\bf 67},(2003),107703.


\bibitem{Mors} P.M.Morse and H.Feshbach, {\it Methods of
Theoretical Physics}, (N.Y.:McGraw-Hill Book Co., 1953).

\bibitem{Frohlich} H.Fr\"ohlich , {\it Adv. Phys. }, {\bf 3},(1954),325.

\bibitem{Alex} C.Alexandrou, R.Rosenfelder and A.W.Schreiber , {\it Phys.Rev. D }, {\bf 62},(2000),085009.

\bibitem{Feynmanpol} R.P.Feynman, {\it Phys. Rev.},
{\bf 97},(1955), 660.

\bibitem{Pekar} S.I.Pekar, {\it Zh. Exper. Teor. Fiz.},
{\bf 16},(1946), 769-774.

\bibitem{Bogol} N.N.Bogoliubov,  {\it Uspekhi Matematicheskih Nauk},
{\bf 2},(1950), 3-24.

\bibitem{Tyablikov} S.V.Tyablikov, {\it Zh. Exper. Teor. Fiz.},
{\bf 21},(1951), 377-388.




\bibitem{Pines} D.Pines, {\it Polarons and Excitons, G.G.Kuper, G.D.Whitfield, eds.}, (N.Y.:Plenum Press,
1962, p. 155).

\bibitem{polaron} {\it Polarons}. Ed. Yu.B.Firsov, (Moscow: Nauka, 1973).


\bibitem{electron} I.D.Feranchuk,  arXiv: hep-th/0309072.


\bibitem{Gross} Gross E.P. {\it Annals of Physics (NY)},
{\bf 99},(1976), 1.



\bibitem{Hainzl} G.Hainzl, M.Lewin and E.Sere , {\it J. of Phys. A}, {\bf 38},(2005), 4483.

\bibitem{OM} I.D.Feranchuk, L.I.Komarov and
A.P.Ulyanenkov, {\it Ann. Phys. (N.Y.)}, {\bf 238},(1995), 370-440;  {\it J. of Phys. A}, {\bf 29},(1996),
4035-4047; I.D.Feranchuk and A.A.Ivanov, {\it\it J. of Phys. A}, {\bf 37},(2004), 9841-9860.

\bibitem{Krylov} L.I.Komarov, E.V.Krylov and I.D.Feranchuk,
{\it Zh. Numerical Math. and Math. Fiz.}, {\bf 18},(1978), 681-691; {\it Zh. Teoret. and Math. Fiz.}, {\bf
32},(1977), 262.


\bibitem{monopol} Ya. M. Shnir {\it Monopole}, (Berlin, Springer, 2005).




\bibitem{Lifshitz} E.M.Lifshitz and L.P.Pitaevskii, {\it Relativistic
Quantum Theory. Part II}, (Moscow, Nauka, 1971).

\bibitem{quasi} N.N.Bogoliubov,  {\it Quasi-mean Values in The Problems
of Statistical Mechanics. In "Selected Works", v.3}, (Kiev, "Navukova Dumka", 1971).








\bibitem{Heitler} W.Heitler, {\it The Quantum Theory of Radiation},
(Oxford, The Clarendon Press, 1954).

\bibitem{Landau} L.D.Landau and E.M.Lifshitz {\it Quantum Mechanics}, (Moscow, Nauka, 1963).

\bibitem{Akhiezer} A.I.Akhiezer and V.B.Beresteckii, {\it Quantum
Electrodynamics}, (Moscow, Nauka, 1969).


\bibitem{Yukalov} V.I.Yukalov,  {\it Vestnik Moscow Univ.(russian)} {\bf 23},(1976), 10;
 {\it Teoret. Matem. Fiz.(russian)} {\bf 28},(1976), 652.

\bibitem{Caswell} W.E.Caswell, {\it Ann. Phys. (N.Y.)},
{\bf 123},(1979), 153.

\bibitem{OM82} I.D.Feranchuk and L.I.Komarov, {\it Phys. Lett. A},
{\bf 88},(1982), 211-213.


\bibitem{Fradkin} E.S.Fradkin, {\it Proceedings of Fiz. Inst.
of Soviet Academy of Science}, {\bf 29},(1965), 1-154; {\it Nucl. Phys.} {\bf 76}, (1966), 588.


\bibitem{NAMN} E.P.Zhidkov, G.I.Makarenko and I.V.Puzynin, {\it Elementary Particles and Atomic Nuclei, Dubna, JINR
}, {\bf 4},(1973), 127-166.

\bibitem{log} S.K.Godunov and V.S.Ryabenkii {\it Introduction to the theory of finite-difference schemes}
(Moscow, Fizmatgiz, 1962).

\bibitem{bipolaron} V.L.Vinetzkii,{\it Zh.of Exper. and Theoret. Physics},{\bf 40},(1961), 1459.

\bibitem{Solodovnikova1}E.P.Solodovnikova and A.N.Tavhelidze,{\it Teoret. and Math. Fiz.},{\bf 21},(1974), 13.


\bibitem{PV} L.D.Landau and S.I.Pekar,{\it Zh.of Exper. and Theoret. Physics},{\bf 18},(1948), 419.

\bibitem{impuls} G. H\"o ler,{\it Zh. Phyzik},{\bf 146},(1956), 372.

\bibitem{Solodovnikova2}E.P.Solodovnikova, A.N.Tavhelidze and O.A.Khrustalov,{\it Teoret. and Math. Fiz.},
{\bf 11},(1972), 317.

\bibitem{TMF} S.T.Zavtrak, L.I.Komarov  and I.D.Feranchuk,{\it Teoret. and Math. Fiz.},{\bf 47},(1981), 55.













\end{thebibliography}
\end{document}